\documentclass[showpacs,amsmath,amssymb]{revtex4}
\usepackage{graphicx}

\begin{document}

\title{Trapping atoms on a transparent permanent-magnet atom chip}

\author{A. Shevchenko\footnote{Email: andrej@focus.hut.fi; Fax.\ +358 9 451
3155}, M. Heili\"o, T. Lindvall, A. Jaakkola, I. Tittonen, and M.
Kaivola} \affiliation{Optics and Molecular Materials, Helsinki
University of Technology, P.O.Box 3500, FI-02015 TKK, Finland}
\author{T. Pfau} \affiliation{5. Physikalisches Institut, Universit\"at Stuttgart, 70550 Stuttgart, Germany}
\date{\today}
\begin{abstract}
We describe experiments on trapping of atoms in microscopic
magneto-optical traps on an optically transparent permanent-magnet
atom chip. The chip is made of magnetically hard ferrite-garnet
material deposited on a dielectric substrate. The confining
magnetic fields are produced by miniature magnetized patterns
recorded in the film by magneto-optical techniques. We trap Rb
atoms on these structures by applying three crossed pairs of
counter-propagating laser beams in the conventional
magneto-optical trapping (MOT) geometry. We demonstrate the
flexibility of the concept in creation and \emph{in-situ}
modification of the trapping geometries through several
experiments.
\end{abstract}
\pacs{03.75.Be, 32.80.Pj, 39.25.+k, 85.70.Ge} \maketitle

Microfabricated devices for trapping and manipulation of ultracold
neutral atoms, known as atom
chips~\cite{Denschlag1999,Folman2000,Cassettari2000}, have been
demonstrated to provide remarkable control of the internal and
external atomic
states~\cite{Treutlein2004,Kruger2003,Hommelhoff2005,Brugger2005,Schumm2005},
and in an essential way to simplify the production of
Bose-Einstein
condensates~\cite{Schumm2005,Hansel2001,Ott2001,Schneider2003,Kasper2003}.
The magnetic-field patterns needed for trapping atoms on these
chips are typically created by driving current through metal wires
that are lithographically fabricated on the chip
surface~\cite{Folman2002}. During the last few years, however,
much attention has been paid to the development of atom chips
based on selectively magnetized permanent
magnets~\cite{Eriksson2004,SinclairF2005,SinclairS2005,Hall2005,Barb2005}.
Such devices, in principle, allow one to get rid of the
electric-power dissipation in the wires, avoid magnetic-field
noise originating from temporal and spatial fluctuations of the
currents, and reduce the near-field noise originating from the
thermal motion of free electrons in the chip. Permanent magnets
also make it possible to create novel geometries for surface
traps, such as storage rings that can serve as miniature rotation
sensors~\cite{Sauer2001,Gupta2005,Arnold2005,Das2002}. In view of
future applications for atom chips, these developments have a
great practical importance.

We introduce a new kind of permanent-magnet atom chip and
demonstrate magneto-optical trapping of $^{85}$Rb atoms on the
surface of this device. The chip is of optically homogeneous
material and transparent to light at near-infrared and infrared
wavelengths. Owing to this property we can make use of the
ordinary MOT geometry with three orthogonal pairs of
counter-propagating laser beams to collect and trap atoms on the
surface instead of using the reflection MOT configuration of the
traditional atom chips. We routinely capture more than $10^6$
atoms in a micro-MOT on a magnetized pattern at a distance of
$\sim100~\mu$m above the chip surface. Being optically
transparent, the device allows unimpeded control and probing of
the on-chip atoms with laser light. Another important feature that
makes our atom chip flexible and simple to operate is the
possibility and ease of \emph{in-situ} reconfiguration of its
trapping potentials. The atoms are trapped above miniature
magnetized patterns which can readily be remotely recorded and
erased by means of conventional magneto-optical recording
techniques, even in the presence of trapped atoms.

To fabricate the chip, a $1.8~\mu$m thick film of magnetically
hard ferrite-garnet, (BiYTmGd)$_3$(FeGa)$_5$O$_{12}$, was grown on
a 500~$\mu$m thick substrate of gadolinium-gallium-garnet
(Gd$_3$Ga$_5$O$_{12}$)~\cite{Jaakkola2005}. The film only absorbs
about 10~\% of the light at $\lambda=780~$nm, which is the
wavelength for trapping Rb atoms. The preferred direction of
magnetization in the film is normal to the surface and the nearly
squared hysteresis loop of the film is characterized by a
saturation magnetization of $\sim20$~mT and coercivity of higher
than 10~mT. We first magnetize the film uniformly and then create
a desired magnetization pattern by locally heating the film with a
scanned, focused cw laser beam at $\lambda=532$~nm, at which
wavelength 80~\% of the light is absorbed in the film. During the
patterning, an external magnetic field of $\sim1$~mT is applied in
the direction opposite to the initial magnetization of the film.
By reversing the direction of the applied magnetic field the
patterns can be erased with the same laser beam. We have created
patterns with dimensions down to the order of $\mu$m with this
technique~\cite{Jaakkola2005}. Typically, 10~mW of power in the
beam is needed to write 10~$\mu$m thick lines on the chip.

\begin{figure}
\centerline{\includegraphics[width=85mm]{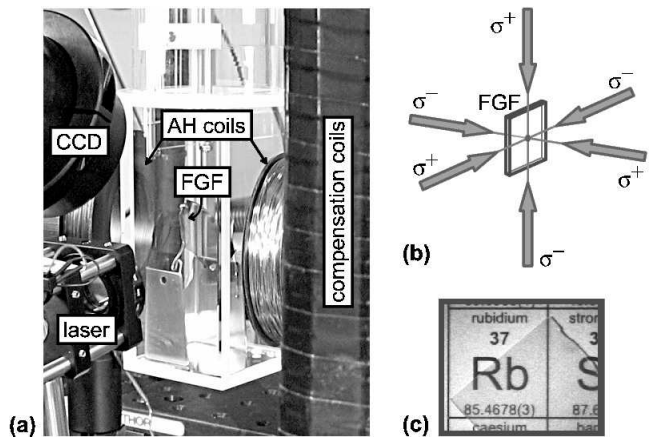}}
\caption{Experimental setup. In (a), CCD - CCD camera, FGF -
ferrite-garnet film on gadolinium-gallium-garnet substrate, laser
- laser head that collimates and $\sigma^+$-polarizes the light
from a fiber. Two anti-Helmholtz coils (AH coils) are positioned
close to the right and left windows of the UHV cell. One pair of
the cooling laser beams, the optical axis of the polarization
imaging system, and the laser beam for magneto-optical patterning
pass through the openings of the coils. (b) Cooling-beam
alignment. (c) Transparency of the atom chip under illumination at
780~nm.}\label{F1}
\end{figure}

The film is placed in a rectangular UHV cell made of fused silica
and connected to a vacuum system that keeps the pressure of
$\lesssim10^{-11}$~mBar (see Fig.~\ref{F1}). The cell is located
in the center of a set of 6 square-shaped magnetic coils of 35~cm
dimension that are used to compensate the background magnetic
fields and to create a uniform external magnetic field on the chip
surface. By controlling the strength and direction of the external
field, a quadrupole field structure can be created at the location
of the magnetized surface pattern. Based on such a magnetic-field
structure, a miniature magneto-optical atom trap at a short
distance from the surface is then created by applying three
orthogonal pairs of retro-reflected laser beams that are
circularly polarized and intersect on the chip surface at the
location of a magnetized pattern. Two of the beam pairs propagate
along the surface, and the third one is let directly through the
chip.

The cooling laser beams are produced by three separate
single-frequency diode lasers that are injection-locked to a
single home-built transmission-grating external-cavity diode
laser~\cite{Merimaa2000}. The external-cavity laser is locked
close to the
$|5^2S_{1/2},F=3\rangle\rightarrow|5^2P_{3/2},F=4\rangle$
transition of $^{85}$Rb. The light from the cooling lasers is
delivered to the setup in three polarization-maintaining optical
fibers. The maximum power in each of the beams is 20~mW, and the
$1/e^2$ diameter of the beams in the cell is about 10~mm. The
light frequency is tuned to the red from the atomic resonance by
one atomic linewidth $\Gamma$. To obtain the repumping radiation,
another laser locked to the
$|5^2S_{1/2},F=2\rangle\rightarrow|5^2P_{3/2},F=2\rangle$
transition is used. This radiation is guided to the cell in the
same fiber as the light of one of the cooling lasers. The power in
the repumping beam is several mW. The trapped atoms are observed
from directions normal and parallel to the film by using two CCD
cameras, SSC-M370CE (Sony) and Pixelfly (PCO).

The atoms are collected to the micro-MOTs from rubidium vapor
evaporated in the cell from a resistively heated Rb dispenser. The
chip is positioned at a distance of 4~cm from the dispenser with
the substrate side facing it. In all the experiments the dispenser
is operated in continuous mode.

\begin{figure}
\centerline{\includegraphics[width=80mm]{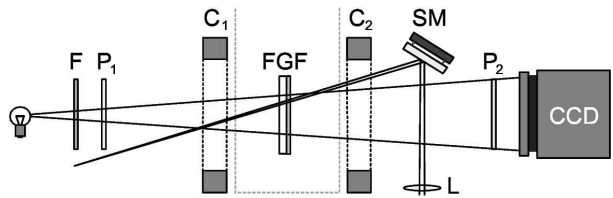}}
\caption{Combined magneto-optical patterning and polarization
imaging system. To record/erase magnetization patterns in the film
(FGF), the electric current in coils C$_1$ and C$_2$ (AH coils in
Fig.~\ref{F1}) is switched on and a cw laser beam
($\lambda=532$~nm) is reflected from a scanning mirror (SM) and
focused with lens (L) onto the film. The imaging system consists
of a CCD camera, a tungsten halogen lamp, a color filter (F), and
two nearly crossed polarizers (P$_1$ and P$_2$).}\label{F2}
\end{figure}

\begin{figure}
\centerline{\includegraphics[width=80mm]{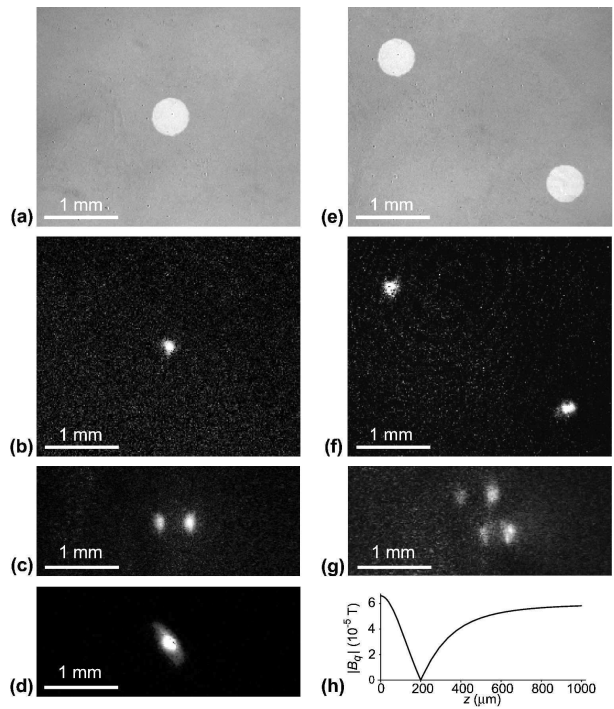}} \caption{Top
row: two different magnetization patterns on the chip. Second row:
front views of the resulting trapped atomic clouds. Third row:
side views of the same traps (in (g) the clouds are located at
different distances from the side-view camera). Bottom row: (d)
front view of an atomic cloud above the pattern (a) when the
capture volume of the trap is increased by applying an additional
large-extent quadrupole field; (h) the absolute value of the
microscopic magnetic quadrupole field as a function of distance
$z$ from the center of spot (a).}\label{F3}
\end{figure}

To record and erase magnetization patterns in the film, a simple
mechanical beam-scanning system was built (see Fig.~\ref{F2}). A
laser beam from a cw laser (Coherent Verdi-V10; $\lambda=532$~nm)
is focused onto the film by reflecting it from a mirror whose tilt
angle is mechanically controlled via a computer. The beam spot
size on the film is adjusted by shifting the lens L along the beam
axis. A mechanical shutter is used to switch the recording/erasing
beam on and off. The external field that defines the magnetization
direction in the recorded pattern is produced by co-running
currents in the AH coils shown in Fig.~\ref{F1}. The setup
includes a polarization-microscopy imaging system for
\emph{in-situ} observation of the recordings. Faraday rotation in
the film is made visible by observing the transmission of thermal
light through the film and two nearly crossed polarizers placed in
front and after the cell~\cite{Jaakkola2005}. An image of the
transmitted-light pattern is recorded with the same camera as is
used to detect the trapped atoms. Figures~\ref{F3}a and e show two
examples of magnetization patterns visualized with this imaging
system.

The magnetic field produced by a given pattern can be calculated
by using the Biot-Savart law~\cite{Jaakkola2005}. For example, the
magnetic field strength at the center of the magnetized circular
spot shown in Fig.~\ref{F3}a is equal to $130~\mu$T. By applying a
uniform magnetic field of $60~\mu$T in the opposite direction one
can obtain a localized quadrupole field above the pattern. The
absolute value of the field, $|B_q|$, calculated as a function of
distance $z$ from the center of the spot along the normal to the
surface is plotted in Fig.~\ref{F3}h. At $z=200~\mu$m, $B_q$ is
zero, and the gradient $\partial B_q/\partial z$ is equal to
$3.5$~mT/cm. The gradients along the $x$ and $y$ directions are
half of this.

For trapping atoms, the intensities of the cooling laser beams are
carefully balanced at the position where the miniature magnetic
quadrupole is created above a selected magnetized pattern. We
first trap atoms in a large magneto-optical trap created with the
aid of two external anti-Helmholtz coils placed close to the cell
windows (see Fig.~\ref{F1}). Then the trap center is shifted
towards the pattern by adjusting the positions of the coils, and
the currents in the coils are gradually decreased to zero.
Simultaneously, the number of atoms in the trap is optimized by
readjusting the retro-reflection angles of the cooling beams.
When eventually the currents in the external coils are switched
off, the atoms remain trapped in the surface trap.
Figure~\ref{F3}b shows the fluorescence image of an atomic cloud
above the pattern of Fig.~\ref{F3}a. In the side view,
Fig.~\ref{F3}c, the cloud is seen together with its reflection
from the surface. The distance of the trap center from the surface
is 200~$\mu$m. The number of atoms in the trap is $7\times10^4$.
In order to increase the number of trapped atoms, we increased the
capture volume of the trap by superimposing on the steep
quadrupole field of the surface trap a weaker quadrupole field of
larger spatial extent produced by the two external anti-Helmholtz
coils outside the cell. The spatial gradient of this field was a
few hundreds of $\mu$T/cm. As a result, the number of atoms in the
trap exceeded $10^6$. This case is illustrated in Fig.~\ref{F3}d.
We then trapped atoms within two identical magnetized spots
positioned at a distance of 3~mm from each other. The spots and
the trapped atomic clouds are shown in Figs.~\ref{F3}e, f, and g.
The additional external quadrupole field was not used in this
case. We note that such micro-MOTs could be created even further
apart from each other if the cooling laser beams were made to have
a larger diameter.

\begin{figure}
\centerline{\includegraphics[width=80mm]{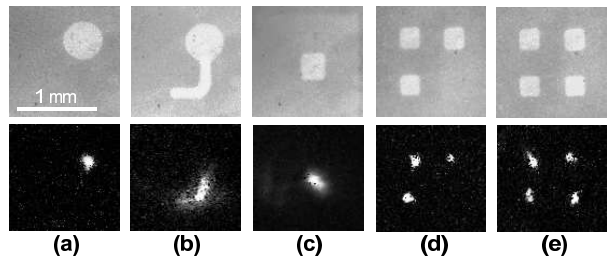}}
\caption{Consecutively recorded and erased magnetization patterns
(top row) and atomic clouds trapped within them (bottom row). The
trap (d) was modified to obtain the trap (e) without significantly
affecting the trapping of the atoms in the three original traps. A
weak auxiliary quadrupole field centered in the middle of the
structure was used to increase the trapping efficiency.}\label{F4}
\end{figure}

In order to demonstrate the \emph{in-situ} reconfigurability of
the trapping potentials, we consecutively created and erased
magnetization patterns within the same area of the film. Some of
the patterns and the atoms trapped above them are shown in
Fig.~\ref{F4}. In Fig.~\ref{F4}b, a curved strip of reversed
magnetization is added to the magnetized spot of Fig.~\ref{F4}a,
and the resulting L-shaped continuation of the MOT is filled with
atoms. This structure was then erased and a square-shaped pattern
was recorded in its place (see Fig.~\ref{F4}c). The number of
atoms in this trap is comparable to that in the trap of
Fig.~\ref{F4}a. We note that, if the new pattern has similar
dimensions and position as the previous one, it is not necessary
to readjust the applied magnetic and optical fields for the new
trap. The trap can be reconfigured even while atoms are confined
in it. The array of micro-MOTs shown in Fig.~\ref{F4}e was
obtained by modifying the array of Fig.~\ref{F4}d without
destroying the functioning of the original micro-traps. The fourth
square was magnetized purely optically, i.e., no additional
writing magnetic field was applied. Even if the value of
magnetization within this additional square is close to zero, the
saturation magnetization of the surrounding area of the film is
high enough for the creation of a micro-MOT. MOT arrays, similar
to those shown in Figs.~\ref{F4}d and e but with somewhat larger
dimensions, have been demonstrated previously by using ribbon UHV
cables and applying the reflection-MOT
principle~\cite{Grabowski2003}.

\begin{figure}
\centerline{\includegraphics[width=80mm]{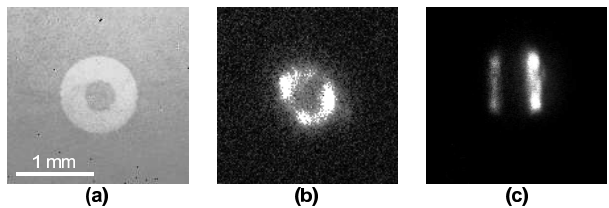}} \caption{A
toroidal trap: (a) magnetization pattern, (b) front view and (c)
side view of trapped atomic cloud.}\label{F5}
\end{figure}

The magneto-optical trapping efficiency turns out to be rather
insensitive to the details of the magnetization-pattern geometry.
It is, for example, possible to collect atoms within some part of
a pattern and then guide the trap along the pattern's structure.
This possibility could provide extra flexibility for the design
and operation of atom-chip circuits. We recorded a toroidal
pattern shown in Fig.~\ref{F5}a that, as a matter of fact, would
be difficult if not impossible to realize using current-carrying
wires. If a uniform magnetic field is applied in the direction
opposite to the magnetization of the torus, the atoms settle in a
ring-shaped MOT, as shown in Figs.~\ref{F5}b and \ref{F5}c. The
slightly uneven distribution of the atoms inside the torus is
mainly explained by the interference and diffraction of the
cooling laser beams at the position of the trap. We believe that
this destructive effect can be substantially reduced by polishing
the film and AR-coating the surfaces of the device.

\begin{figure}
\centerline{\includegraphics[width=80mm]{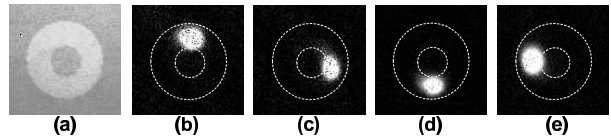}} \caption{A surface
MOT translated along a toroidal magnetization pattern: (a)
magnetization pattern, (b)-(e) a series of images of the trapped
atomic cloud.}\label{F6}
\end{figure}

By periodically modulating the $x$- and $y$-components of a weak
external magnetic field of the compensation coils with a mutual
phase difference of $\pi/2$, we could drive the trap center into
circular motion along the torus. The modulation is accomplished by
modulating the currents in the 35~cm compensation coils.
Figures~\ref{F6}b-e show a sequence of images of the trapped atoms
separated in time by a quarter of the modulation period. In this
case, an auxiliary quadrupole field, with a spatial gradient of
300~$\mu$T/cm, was added to increase the number of atoms. The
modulation amplitudes were $27~\mu$T.

In conclusion, we have demonstrated magneto-optical trapping of
atoms on a transparent permanent-magnet atom chip. This novel
approach to the creation of atom chips can provide several
advantages over conventional techniques based on current-carrying
wires and over the other up-to-date techniques employing permanent
magnets. Our traps are readily reconfigurable \emph{in situ}.
Essentially free-format trap patterns can be realized, as
demonstrated by the example of a ring-shaped trap. Ring-shaped
traps are particularly interesting due to the possibility of
applying them in a Sagnac-type atom
interferometer~\cite{Sauer2001,Gupta2005,Arnold2005,Das2002}. The
device is transparent to light, which provides unimpeded control
of atoms with laser radiation. In particular, an ordinary MOT
instead of a reflection MOT geometry was used to collect atoms
close to the surface. Microscopic magneto-optical traps described
in this work are formed at a distance of a few $100~\mu$m from the
surface and they contain more than $10^6$ atoms. We also created
surface-mounted arrays of micro-MOTs. Such an array can be used to
prepare multiple atomic samples on the chip. Each of these samples
can then be processed individually, by using, e.g., a nearly
resonant focused optical field.

There are no electric currents applied to the device.
Consequently, there is no electric-power dissipation nor temporal
or spatio-temporal current fluctuation. Since the device is made
of dielectric material, magnetic-field noise due to thermal
electrons is insignificant at short distances from the
surface~\cite{Henkel2003}. Purely magnetic micro-traps with a trap
depth of up to 1~mK are realizable on an atom chip of this
type~\cite{Jaakkola2005}. Such traps can be loaded with atoms from
surface-mounted micro-MOTs described in the present work and be
used for Bose-Einstein condensation and for experiments on atom
interferometry.

We acknowledge financial support from the Academy of Finland and
the Jenny and Antti Wihuri Foundation and thank E.I.~Il'yashenko
and T.H.~Johansen, University of Oslo, for fruitful collaboration.

\end{document}